# On the Complexity of Solving a Bivariate Polynomial System

Pavel Emeliyanenko and Michael Sagraloff

Max-Planck-Institut für Informatik, Saarbrücken, Germany

**Abstract.** We study the complexity of computing the real solutions of a bivariate polynomial system using the recently proposed algorithm BISOLVE [3]. BISOLVE is a classical elimination method which first projects the solutions of a system onto the $x$- and $y$-axes and, then, selects the actual solutions from the so induced candidate set. However, unlike similar algorithms, BISOLVE requires no genericity assumption on the input nor it needs any change of the coordinate system. Furthermore, extensive benchmarks from [3] confirm that the algorithm outperforms state of the art approaches by a large factor. In this work, we show that, for two polynomials $f, g \in \mathbb{Z}[x,y]$ of total degree at most $n$ with integer coefficients bounded by $2^\tau$, BISOLVE computes isolating boxes for all real solutions of the system $f = g = 0$ using $\tilde{\mathcal{O}}(n^8\tau^2)$ bit operations[1], thereby improving the previous record bound by a factor of at least $n^2$.

## 1 Introduction

Systems of polynomial equations naturally arise in many fields of science and engineering. In computational geometry and computer graphics, there is a particular interest in the study of polynomial systems in two or three variables. For instance, all existing exact and complete algorithms for computing the topology or arrangements of algebraic curves [6,13] (and surfaces [4]) are crucially based on determining the critical points, which are in turn the solutions of a bivariate polynomial system. In this work, we investigate in the bit complexity analysis of the recently proposed algorithm BISOLVE [3] to isolate the real solutions of a bivariate polynomial system

$$f(x,y) = \sum_{i,j \in \mathbb{N}: i+j \leq n} f_{ij} x^i y^j = 0, \quad g(x,y) = \sum_{i,j \in \mathbb{N}: i+j \leq n} g_{ij} x^i y^j = 0, \quad (1.1)$$

where $f, g \in \mathbb{Z}[x,y]$ are polynomials of total degree at most $n$ and with integer coefficients bounded by $2^\tau$, $\tau \in \mathbb{N}$. For short, we will also write that $f$ and $g$ have *magnitude* $(n, \tau)$. Henceforth, we assume that $f$ and $g$ share no common non-trivial factor in $\mathbb{Z}[x,y] \setminus \mathbb{Z}$ which, due to Bézout's Theorem, is equivalent to the existence of finitely many (complex) solutions of (1.1). BISOLVE computes a set of disjoint boxes $B_k \subset \mathbb{R}^2$ such that the union of all $B_k$ contains

$$V_\mathbb{R} := \{(x,y) \in \mathbb{R}^2 | f(x,y) = g(x,y) = 0\},$$

---

[1] $\tilde{\mathcal{O}}$ indicates that polylogarithmic factors in $\tau$ and $n$ are omitted.

the set of all *real* solutions of (1.1), and each $B_k$ is *isolating*, that is, it contains exactly one solution. We show that to achieve the latter task, BISOLVE demands for $\tilde{\mathcal{O}}(n^8\tau^2)$ bit operations, improving the previous record bound by a factor of $n^2$ or more (depending on whether $\tau$ or $n$ is dominating). Our analysis uses two recent results on isolating [17] and refining [14] the real roots of a univariate polynomial. Yet, we remark that the so obtained complexity bound for BISOLVE is *not* merely due to the improved complexity bounds for the root isolation and refinement step, but also due to the effectiveness of the novel inclusion predicate used in the *validation step* of BISOLVE to certify or discard candidate solutions. We remark that, in all previous algorithms to compute $V_\mathbb{R}$, the computation of the common roots of $f(\alpha, y)$ and $g(\alpha, y)$ was achieved by considering a corresponding signed remainder sequence, with $\alpha$ the projection of a solution in $\mathbb{R}^2$ onto the real axis. In practice, the latter computation often turns out to be a bottleneck of the overall approach. BISOLVE does not follow this approach which has proven to be favorable in practice. Our analysis also shows that the final validation step is less hard than isolating the roots of the resultant polynomial, which seems to be the bottleneck. In addition, we would like to emphasize that BISOLVE is not a "galactic" algorithm just designed to achieve good complexity bounds. In contrast, the high performance of BISOLVE in practice is confirmed by many experiments; we refer to [3] for an extensive comparison of the algorithm with other state of the art approaches such as LGP [7] or Maple's ISOLATE.

*Related work.* An early result on the complexity analysis appears in [12], where the closely related problem of computing the topology of an algebraic curve is considered. The authors analyze the algorithm TOP and derive a complexity bound of $\tilde{\mathcal{O}}(N^{14})$ bit operations, with $N = \max(n, \tau)$. Another work [8] presents several methods to solve a bivariate polynomial system. The first method, GRID, first projects the solutions onto orthogonal axes and, then, matches them by means of a SIGN_AT procedure which uses evaluation of a signed remainder sequence of $f$ and $g$. The complexity of GRID is bounded by $\tilde{\mathcal{O}}(N^{14})$ bit operations, where the overall cost is dominated by the SIGN_AT operations. The second approach, M_RUR, is based on subresultants and RUR (rational univariate representation) techniques and achieves a bit complexity $\tilde{\mathcal{O}}(n^{10}(n^2+\tau^2)) = \tilde{\mathcal{O}}(N^{12})$. The third approach, G_RUR, achieves the same complexity as M_RUR but relies on the computation of the GCD of the square-free parts of $f(\alpha, y)$ and $g(\alpha, y)$. It should not be revealed that, using the improved complexity bounds for the univariate root isolation, one may also obtain improvements of the corresponding complexity bounds for M_RUR (only in a sheared system) and G_RUR. However, it seems that the so obtained bounds are still weaker than the bound achieved by BISOLVE. Namely, the dependence on $n$ in the final steps of M_RUR and G_RUR is considerably larger (i.e., by a factor $n^2$ when $n$ is dominating). We further remark that the complexity analysis of G_RUR is based on the modular algorithm in [19] whose complexity is only given with respect to *average running time* (cf. Section 3.2 in [19]). In his dissertation, M. Kerber describes randomized algorithms to analyze the topology of a single algebraic curve and to compute arrangements of such curves. A detailed analysis of the "curve-pair analysis" which solves the subproblem of finding the solutions of a bivariate system shows that the corresponding bit complexity is bounded by an *expected number* of $\tilde{\mathcal{O}}(n^{10}(n+\tau)^2)$ bit operations; see [13, Section 3.3.4].

*Outline.* Section 2 introduces some notations which are used throughout the argument. In Section 3, we briefly review the algorithm BISOLVE omitting some technical details and filtering techniques to keep the presentation simple. The complexity analysis is given in Section 4. We analyze the three main steps of the algorithm separately and, then, combine the results yielding the overall complexity. Finally, in Section 5, we give some concluding remarks.

## 2 Setting

We express the input polynomials $f$ and $g$ in (1.1) as *univariate* polynomials in $x$ and $y$ of degrees $n_x$ and $n_y$, respectively:

$$f(x,y) = \sum_{i=0}^{n_x} f_i^{(x)}(y) x^i = \sum_{i=0}^{n_y} f_i^{(y)}(x) y^i, \quad g(x,y) = \sum_{i=0}^{n_x} g_i^{(x)}(y) x^i = \sum_{i=0}^{n_y} g_i^{(y)}(x) y^i,$$

where $f_i^{(y)}, g_i^{(y)} \in \mathbb{Z}[x]$ and $f_i^{(x)}, g_i^{(x)} \in \mathbb{Z}[y]$. Throughout the paper, it is assumed that $n_x, n_y \leq n$. We denote the *Sylvester's matrix* associated with $f$ and $g$ by $S^{(y)} = S^{(y)}(f,g)$ whose entries are the coefficients $\{f_i^{(y)}\}$ and $\{g_i^{(y)}\}$; see [11, p. 286] for the definition. The resultant $R^{(y)} = \mathrm{res}(f,g,y) \in \mathbb{Z}[x]$ of $f$ and $g$ with respect to the variable $y$ is the determinant of $S^{(y)}$. By analogy, $R^{(x)} = \mathrm{res}(f,g,x)$ defines the resultant with respect to the variable $x$, and $S^{(x)}(f,g)$ is the associated Sylvester's matrix with entries $\{f_i^{(x)}\}$ and $\{g_i^{(x)}\}$. If this causes no ambiguity, we also write $R$ omitting the variable index and by $R^*$ the square-free part of $R$. For an interval $I = (a,b) \subset \mathbb{R}$, $w_I := b-a$ denotes the *width*, $m_I := (a+b)/2$ the *center* and $r_I := (b-a)/2$ the *radius* of $I$. A disc in $\mathbb{C}$ is denoted by $\Delta = \Delta_r(m)$, where $m \in \mathbb{C}$ defines the center of $\Delta$ and $r \in \mathbb{R}^+$ its radius. For a (not necessarily square-free) polynomial $F(x) = \sum_{i=0}^{n} F_i x^i \in \mathbb{R}[x]$ with distinct roots $z_1 \ldots z_m \in \mathbb{C}$, the *separation* $\sigma_i := \sigma(z_i, F)$ of a root $z_i$ is defined as the minimal distance of $z_i$ to any root $z_j \neq z_i$. The root separation $\sigma(F)$ of $F$ is defined as the minimum of all $\sigma_i$, and $\Sigma(F) = \sum_{i=1}^{m} \log \sigma(z_i, F)^{-1}$. We finally denote $\Gamma(F) := \max_i |z_i|$ the maximal absolute value of all $z_i$.

## 3 Review of the algorithm

In this section, we recall the algorithm BISOLVE to make the paper self-contained; for further details and filtering techniques used in the actual realization, we refer to [3]. At the highest level, BISOLVE comprises three subroutines which we consider separately.

PROJECT: We begin with projecting the complex solutions of (1.1) onto the $x$- and $y$-axes. In other words, we consider the two sets:

$$V_{\mathbb{C}}^{(x)} := \{x \in \mathbb{C} | \exists y \in \mathbb{C} \wedge f(x,y) = g(x,y) = 0\}$$
$$V_{\mathbb{C}}^{(y)} := \{y \in \mathbb{C} | \exists x \in \mathbb{C} \wedge f(x,y) = g(x,y) = 0\}$$

and compute their restrictions $V_{\mathbb{R}}^{(x)} := V_{\mathbb{C}}^{(x)} \cap \mathbb{R}$ and $V_{\mathbb{R}}^{(y)} := V_{\mathbb{C}}^{(y)} \cap \mathbb{R}$ to the real values. The real solutions $V_{\mathbb{R}}$ of (1.1) are then contained in the product

$$\mathcal{C} := V_{\mathbb{R}}^{(x)} \times V_{\mathbb{R}}^{(y)} \subset \mathbb{R}^2, \tag{3.1}$$

which we denote the set of *candidate solutions* for (1.1). To obtain the projection sets $V_{\mathbb{R}}^{(x)}$ and $V_{\mathbb{R}}^{(y)}$, we first compute the resultants $R^{(y)}$ and $R^{(x)}$, respectively, and extract the square-free part $R^*$ of either polynomial ($R = R^{(y)}$ or $R = R^{(x)}$ for short). Next, we isolate the real roots $\alpha_i$ of $R^*$ using RIsolate, a recently proposed Descartes method [17] mentioned in the introduction. RIsolate is an exact method which works with approximations of the coefficients of a polynomial and achieves a significantly better bit complexity compared to the classical Descartes algorithm [2, Remark 10.52]. Moreover, in contrast to asymptotically fast algorithms such as those proposed by Pan or Schönhage (see [16] for an overview), RIsolate is practical and easy to implement; see [18] for an implementation within Mathematica. The isolating intervals $I := I(\alpha)$ returned by RIsolate satisfy

$$\sigma(\alpha, R)/(4 \deg R^*)^2 < w_I < 2(\deg R^*) \cdot \sigma(\alpha, R), \tag{3.2}$$

which will be important for us in the context of our complexity analysis. This concludes the first step of the algorithm.

SEPARATE: In this step, the real roots of $R$ are further separated from the complex ones. For each root $\alpha$, we refine a corresponding isolating interval $I := I(\alpha)$ such that the disc $\Delta_{8r_I}(m_I)$ does not contain any root of $R$ except $\alpha$. RIsolate returns an interval $I$ which fulfills the inequality in (3.2). Thus, after performing $\lceil \log(16 \deg(R^*)) \rceil$ bisection steps to refine $I$, the disc $\Delta_{8r_I}(m_I)$ isolates $\alpha$ as well. Then, we compute

$$LB(\alpha) := 2^{-2 \deg R} |R(m_I - 2r_I)|, \tag{3.3}$$

which constitutes a lower bound for $|R(x)|$ on the boundary of $\Delta(\alpha) := \Delta_{2r_I}(m_I)$, that is, $|R(x)| > LB(\alpha)$ for all $x \in \partial \Delta(\alpha)$; see [3, Thm. 3.2] for a proof. We remark that in the initial description of the algorithm (as well as in the realization), we consider a different predicate to ensure that $\Delta_{8r_I}(m_I)$ is isolating. The latter predicate is based on a local Taylor expansion of $R^*$ at $m_I$, does not require (3.2) and shows a slightly more adaptive behavior in practice. It is rather straight forward to prove that also using this method the overall approach achieves the claimed bit complexity bound. However, for the sake of simplicity, we decided to exploit the inequality in (3.2) for our complexity analysis.

To sum up, at the end of SEPARATE, for each real root $\alpha$ of $R^{(y)}$ (and $\beta$ of $R^{(x)}$), we have an isolating interval $I(\alpha)$ (and $I(\beta)$) and an isolating disc $\Delta(\alpha) := \Delta_{2r_{I(\alpha)}}(m_{I(\alpha)})$ (and $\Delta(\beta)$). Then, each real solution of system (1.1) is contained in a polydisc $\Delta(\alpha, \beta) := \Delta(\alpha) \times \Delta(\beta) \subset \mathbb{C}^2$, and each of these polydiscs contains at most one solution. In addition, for each point $(x, y)$ on the boundary of a polydisc $\Delta(\alpha, \beta)$, we have $|R^{(y)}(x)| > LB(\alpha)$ or $|R^{(x)}(y)| > LB(\beta)$.

VALIDATE: The goal of this final stage is to determine all candidates $(\alpha, \beta)$ which are actually solutions of (1.1) and to exclude the remaining ones. Again, to facilitate the complexity analysis, we assume that the actual solutions are chosen exclusively based on the *inclusion test* outlined below. We remark that the efficiency of the actual implementation is further due to a series of filtering techniques to rapidly exclude the majority of candidates such as tests based on interval arithmetic or a bitstream Descartes algorithm [10].

In SEPARATE, we have already computed lower bounds $LB(\alpha)$ and $LB(\beta)$ for the values of $|R^{(y)}|$ and $|R^{(x)}|$ at the boundaries of $\Delta(\alpha)$ and $\Delta(\beta)$, respectively. We now rewrite $R^{(y)}$ in terms of cofactors $u^{(y)}$ and $v^{(y)}$ (see [11, p. 287]):

$$R^{(y)}(x) = u^{(y)}(x,y)f(x,y) + v^{(y)}(x,y)g(x,y) \qquad (3.4)$$

where $u^{(y)}$ and $v^{(y)}$ are determinants of "Sylvester-like" matrices $U^{(y)}$ and $V^{(y)}$. These matrices are obtained from $S^{(y)}(f, g)$ by replacing the last column with vectors $(y^{n_y-1} \ldots y\ 1\ 0 \ldots 0)^T$ and $(0 \ldots 0\ y^{n_y-1} \ldots y\ 1)^T$ of appropriate size, respectively. Next, we compute upper bounds $UB(\alpha, \beta, u^{(y)})$ and $UB(\alpha, \beta, v^{(y)})$ for $|u^{(y)}|$ and $|v^{(y)}|$ on $\Delta(\alpha, \beta)$, respectively. These bounds can be obtained by bounding the absolute values of the entries in $U^{(y)}$ and $V^{(y)}$ and then applying Hadamard's inequality to $U^{(y)}$ and $V^{(y)}$. We remark that the computation of the latter bounds is done *without* explicitly computing the cofactors (which are typically very large expressions). Cofactor polynomials $u^{(x)}$, $v^{(x)}$ and respective upper bounds $UB(\alpha, \beta, u^{(x)})$, $UB(\alpha, \beta, v^{(x)})$ are defined in an analogous way for the resultant polynomial $R^{(x)}$. The inclusion test based on a homotopy argument is now formulated as follows (see [3, Thm. 4] for a proof):

**Theorem 1.** *If there exists an $(x_0, y_0) \in \Delta(\alpha, \beta)$ with*

$$UB(\alpha, \beta, u^{(y)}) \cdot |f(x_0, y_0)| + UB(\alpha, \beta, v^{(y)}) \cdot |g(x_0, y_0)| < LB(\alpha), \qquad (3.5)$$

$$UB(\alpha, \beta, u^{(x)}) \cdot |f(x_0, y_0)| + UB(\alpha, \beta, v^{(x)}) \cdot |g(x_0, y_0)| < LB(\beta), \qquad (3.6)$$

*then $\Delta(\alpha, \beta)$ contains a solution of (1.1) and, thus, $f(\alpha, \beta) = 0$.*

All candidate solutions $(\alpha, \beta) \in \mathcal{C}$ are now treated as follows: Let $B(\alpha, \beta) = I(\alpha) \times I(\beta) \subset \mathbb{R}^2$ be the corresponding *candidate box*. Each candidate box is then refined with the quadratic interval refinement (QIR for short) method from [14] until either we can ensure that $f(\alpha, \beta) \neq 0$ or $g(\alpha, \beta) \neq 0$ based on interval arithmetic on $B(\alpha, \beta)$ or, for an arbitrary point $(x_0, y_0) \in B(\alpha, \beta)$, the inequalities (3.5) and (3.6) are fulfilled. In the latter case, Theorem 1 guarantees that $(\alpha, \beta)$ is a solution of (1.1). We refer to Section 4.3 for the details of the evaluation using interval arithmetic.

# 4 Complexity analysis

Throughout the analysis, we assume that the multiplication of two integers is always done in *asymptotically fast* way. In other words, the bit complexity to multiply two $k$-bit integers is assumed to be $M(k) = \mathcal{O}(k \log k \log \log k) = \tilde{\mathcal{O}}(k)$.

## 4.1 Project

Computing the resultant polynomials $R = R^{(x)}$ (and $R = R^{(y)}$) using the subresultant PRS algorithm needs $\tilde{\mathcal{O}}(n^7(\log n + \tau))$ bit operations. $R$ has magnitude

$$(n^2, \mathcal{O}(n(\log n + \tau))), \tag{4.1}$$

see [13, Thms. 2.4.14-17], where $R = \mathrm{Sres}_0(f, g, y)$, the 0-th subresultant of $f$ and $g$ with respect to $y$. Next, we compute the square-free part $R^*$ of $R$. This can be done with $\tilde{\mathcal{O}}(n^5(\tau + \log n))$ bit operations (see [13, Thm. 2.4.21]), and $R^*$ is of magnitude

$$(n^2, \mathcal{O}(n(n + \tau))). \tag{4.2}$$

Finally, the real roots of $R^*$ are isolated using $\mathbb{R}$ISOLATE as outlined in Section 3. The complexity of the root isolation is summarized below (see [17, Thm. 18]):

**Theorem 2.** *Let $F$ be a square-free polynomial of magnitude $(d, \mu)$. Then, isolating the real roots of $F$ demands for $\tilde{\mathcal{O}}(d(\Sigma(F) + d\log \Gamma(F))^2)$ bit operations, with $\Sigma(F)$ and $\Gamma(F)$ as defined in Section 2.*

We aim to apply Theorem 2 to $F := R^*$. Obviously, $\Gamma(R^*) = \Gamma(R)$ and $\Gamma(R^*) = \Gamma(R)$, and a bound on $\Sigma(R)$ is provided by the following theorem:

**Theorem 3.** *For a (not necessarily square-free) polynomial $F(x)$ with integer coefficients of magnitude $(d, \mu)$ and distinct complex roots $z_1, \ldots, z_r$ $(r \leq d)$, it holds that $\Sigma(F) = \mathcal{O}(d\mu \log(d\mu))$.*

*Proof.* The proof essentially follows the same lines as the proof in [17, Appendix 6.2], with the exception that we use the Davenport-Mahler bound for non square-free polynomials. We refer to Appendix A for a full argument.

Now, by Theorem 3 and (4.1), it follows that:

$$\Sigma(R^*) = \mathcal{O}(n^2 \cdot n(\tau + \log n) \cdot \log(n^3\tau)) = \tilde{\mathcal{O}}(n^3\tau). \tag{4.3}$$

Observe that, due to Cauchy's bound [2, Corollary 10.4], the absolute values of the roots of $R$ (and $R^*$) are bounded by a

$$B = 2^{\mathcal{O}(n(\log n + \tau))}. \tag{4.4}$$

Hence, from (4.3), (4.4) and Theorem 2, we conclude that the complexity of isolating the real roots of $R^*$ is bounded by:

$$\tilde{\mathcal{O}}(n^2(\Sigma(R^*) + n^2 \log B)^2) = \tilde{\mathcal{O}}(n^2(n^3\tau + n^3\tau)^2) = \tilde{\mathcal{O}}(n^8\tau^2), \tag{4.5}$$

which determines the complexity of PROJECT.

### 4.2 Separate

Let $\alpha$ be a real root of one of the resultant polynomials $R$ and $I = I(\alpha)$ a corresponding isolating interval returned by $\mathbb{R}$Isolate. In our description of Separate, we have already seen that $I = I(\alpha)$ is refined by means of $\mathcal{O}(\log n)$ bisection steps to guarantee that the disc $\Delta_{8r_I}(m_I)$ isolates $\alpha$. The bit complexity of one bisection step for a polynomial $F$ of magnitude $(d, \mu)$ and an interval of bitsize $\theta$ is $\tilde{\mathcal{O}}(d(\mu + d\theta))$ because this is essentially evaluation of $F$ at the endpoints of $I$; see [13, Prop. 2.5.1].[2] The maximal bitsize of the interval $I = I(\alpha)$ is essentially determined by the separation of $\alpha$. Namely, $I$ is refined to a width larger than $\sigma(\alpha, R)/(1024n^6)$ because we started with an interval of width larger than $\sigma(\alpha, R)/(16n^4)$ (see (3.2) for a bound on the size of the initial isolating interval) and then performed less than $32n^2$ further bisection steps. Hence, we obtain $\theta(\alpha) = \mathcal{O}(\log B + \log(1/\sigma(\alpha, R)) + \log n)$ for the bitsize of $I(\alpha)$, where $B$ is the root bound for $R$, see (4.4). It follows that the complexity to refine the isolating intervals for the real roots $z_1, \ldots, z_k$ of $R^*$ in Separate is bounded by $((d, \mu) = (n^2, \tilde{\mathcal{O}}(n(n + \tau))))$ a bound on the magnitude of $R^*$:

$$\sum_{i=1}^{k} \tilde{\mathcal{O}}(d(\mu + d\theta(z_i))) = \sum_{i=1}^{k} \tilde{\mathcal{O}}(n^5\tau - n^4 \log \sigma(z_i, R)) = \tilde{\mathcal{O}}(n^7\tau), \qquad (4.6)$$

where we use that $-\sum_{i=1}^{k} \log \sigma(z_i, R) < \Sigma(R^*) + n^2 \log(2B) = \tilde{\mathcal{O}}(n^3\tau)$ because each root $z_j$ of $R^*$ not which is not considered in the left sum has separation less than $2B = \tilde{\mathcal{O}}(n\tau)$ according to (4.4) and $\Sigma(R^*) = \tilde{\mathcal{O}}(n^3\tau)$ according to (4.3).

### 4.3 Validate

**Estimating lower and upper bounds** In the final stage, we have a set of candidate solutions $\mathcal{C}$ and corresponding disjoint polydiscs $\Delta(\alpha, \beta) := \Delta(\alpha) \times \Delta(\beta) \subset \mathbb{C}^2$. Each of the polydiscs contains at most one solution of (1.1) (namely, $(\alpha, \beta)$). It remains to determine the actual solutions based on the inclusion test from Theorem 1 and to exclude the other candidates by means of interval arithmetic. We split the complexity analysis in two parts. In the first part, we derive a lower bound for the value $LB(\alpha)$ as well as an upper bound for the values $UB(\alpha, \beta, u^{(y)})$ and $UB(\alpha, \beta, v^{(y)})$ needed by the inclusion predicate. In the second part, we estimate how good each candidate $(\alpha, \beta)$ needs to be approximated in order to certify or it as a solution or to discard it.

*Estimating the lower bounds.* In this section, we derive a lower bound for $LB(\alpha)$ which is in turn a lower bound for $|R^{(y)}|$ on the boundary of $\Delta(\alpha)$; see (3.3) for the definition of $LB(\alpha)$. By analogy, we obtain a similar bound also for $LB(\beta)$, the lower bound for $|R^{(x)}|$ on the boundary of $\Delta(\beta)$. Remark that the polynomial $R = R^{(y)}$ might have multiple roots. The key idea to keep the bounds tight is to exploit the fact that the separation of a root actually depends on its multiplicity. We begin with the following auxiliary lemmas.

---

[2] The bitsize of an interval is defined as maximal bitsize of its rational endpoints.

**Lemma 1.** *[1, Thm. AB and Thm. 1] Let $F$ be an integer polynomial of degree $d \geq 2$. Suppose that $\alpha$ is a zero of $F(x)$ of order $s$ and $\beta$ a zero of $F(x)$ of order $t$. If $\alpha \neq \beta$ and $t \geq s$, then*

$$|\alpha - \beta| \geq 2^{-d/t}(d+1)^{-d/t - 3d/(2st)}|F|_\infty^{-2d/(st)} \max\{1, |\alpha|\} \max\{1, |\beta|\}.$$

**Lemma 2.** *[5, Lem. A.7]. Let $F$ be a non-constant integer polynomial of degree $d$. Let $\xi$ be a complex number and $\alpha$ be the root of $F(x)$ which is closest to $\xi$ (i.e. $|\xi - \alpha|$ is minimal for all roots $\alpha$ of $F$). Then, denoting by $s$ the multiplicity of $\alpha$ as root of $F(x)$, we have:*

$$|\xi - \alpha|^s \leq d^{d+3d/(2s)}|F|_\infty^{2(d/s-1)}|F(\xi)|.$$

Suppose $\alpha = z_i$ is a root with multiplicity $m_i$ of $R$ and let $I = I(z_i)$ be the corresponding isolating interval for $z_i$ obtained in SEPARATE. In Section 4.2, we have already shown that $r_I \geq \sigma(\alpha, R)/(1024n^6)$ and, thus, the distance of $z := m_I - 2r_I$ to $z_i$ is larger than $\sigma(\alpha, R)/(1024n^6)$. Let $z_j \neq z_i$ be another root of $R$ closest to $z_i$ (i.e., $|z_i - z_j| = \sigma(z_i, R)$) with multiplicity $m_j$. Then, by Lemma 1, it follows, with $m = \max(m_i, m_j)$, that

$$\begin{aligned}\log|z_i - z_j| &\geq -n^2/m - (n^2/m + 3n^2/(2m_i m_j))\log(n^2+1) - 2n^2/(m_i m_j)\log|R|_\infty \\ &> -16n^2 \log n/m - n^2/(m_i m_j)\log|R|_\infty = -\tilde{\mathcal{O}}(n^3\tau/m),\end{aligned} \tag{4.7}$$

where we used that $\log|R|_\infty = \mathcal{O}(\log n \cdot n(\log n + \tau))$. By the construction of $I$, $z_i$ is the closest root to $z$. Hence, Lemma 2 yields

$$\begin{aligned}\log|R(z)| &\geq m_i \log|z - z_i| - 2(n^2/m_i - 1)\log|R|_\infty - 2n^2(1 + 3/(2m_i))\log n \\ &\geq m_i \log(|z_i - z_j|/(1024n^6)) - (2n^2/m_i)\cdot \log|R|_\infty - 5n^2 \log n.\end{aligned} \tag{4.8}$$

By combining (4.7) and (4.8), we obtain

$$\log LB(\alpha) > \log|R(z)| - 2n^2 = -\tilde{\mathcal{O}}(m_i \cdot n^3\tau/m_i + n^3\tau/m_i + n^2) = -\tilde{\mathcal{O}}(n^3\tau). \tag{4.9}$$

In completely analogous manner, we show that $\log LB(\beta) > -\tilde{\mathcal{O}}(n^3\tau)$.

*Estimating the upper bounds.* We now investigate in estimating the upper bounds $UB(\alpha, \beta, u^{(y)})$ and $UB(\alpha, \beta, v^{(y)})$ for $|u^{(y)}|$ and $|v^{(y)}|$ on $\Delta(\alpha, \beta)$. In order to do so, we apply Hadamard's inequality to the matrices $U^{(y)}$ and $V^{(y)}$, see Section 3. By analogy, these estimates then also extend to the upper bounds $UB(\alpha, \beta, u^{(x)})$ and $UB(\alpha, \beta, v^{(x)})$ for $|u^{(x)}|$ and $|v^{(x)}|$ on $\Delta(\alpha, \beta)$.

In the realization, to compute the upper bounds on the cofactors, we first use interval arithmetic on a box in $\mathbb{C}^2$ to estimate the absolute values of respective matrix entries $U_{ij}$ and $V_{ij}$ on $\Delta(\alpha, \beta)$, and then apply Hadamard's bound. In the complexity analysis, we follow a slightly different but even simpler approach: From the construction of $\Delta(\alpha, \beta)$, the disc $\Delta(\alpha)$ has radius less than

$\sigma(\alpha, R^{(y)})/4$, and the disc $\Delta(\beta)$ has radius less than $\sigma(\beta, R^{(x)})/4$; see Section 3 for details. Hence, it is clear that both radii are bounded by $B = 2^{O(n(\tau + \log n))}$, where $B$ is the upper bound from (4.4) for the modulus of the roots of $R^{(x)}$ and $R^{(y)}$. It follows that, for each point $(x, y) \in \Delta(\alpha, \beta)$, we have $|x|, |y| < 2B$. Recall that, the entries of $U^{(y)}$ are the coefficients $f_i^{(y)}(x)$ and $g_i^{(y)}(x)$ which are polynomials of degree $n_x \leq n$ and with integer coefficients of bitsize $\tau$ or less. Thus, for $(x, y) \in \Delta(\alpha, \beta)$, we can bound them as follows:

$$|f_i^{(y)}(x)| \leq \sum_{j=0}^{n_x} |f_{ij}^{(y)}|(2B)^j \leq (n_x + 1) \cdot |f_{ij}^{(y)}|(2B)^{n_x} \leq 2^{n+1} n \cdot \tau B^n,$$

and a corresponding bound holds for $|g_i^{(y)}(x)|$ as well. Hence, it follows that $\log |U_{ij}^{(y)}(x, y)| = \mathcal{O}(n \log B + \tau + n) = \mathcal{O}(n^2(\log n + \tau))$. Now, by Hadamard's inequality $|u^{(y)}| = |\det(U^{(y)})| < \prod_i \|U_i^{(y)}\|_2$, where $\|U_i^{(y)}\|_2$ is the 2-norm of the $i$-th row vector of $U^{(y)}$. Since the absolute value of each entry of $U^{(y)}$ is bounded by $2^{\mathcal{O}(n^2(\log n + \tau))}$, we have $\log \|U_i^{(y)}\|_2 = \tilde{\mathcal{O}}(n^2 \tau)$. Taking the product of all bounds on the 2-norms yields $\log |u^{(y)}| = \tilde{\mathcal{O}}(n^3 \tau)$ and, thus,

$$UB(\alpha, \beta, u^{(y)}) = \tilde{\mathcal{O}}(n^3 \tau)). \tag{4.10}$$

This bound is also valid for $UB(\alpha, \beta, v^{(y)})$, $UB(\alpha, \beta, u^{(x)})$ and $UB(\alpha, \beta, v^{(x)})$.

**The inclusion test** Let us first assume that a candidate box $\mathcal{B} := B(\alpha, \beta) = I(\alpha) \times I(\beta) \subset \mathbb{R}^2$ contains a real solution of (1.1). Using our bounds from (4.9) and (4.10), the inequalities in (3.5) and (3.6) rewrite as:

$$|f(x_0, y_0)| + |g(x_0, y_0)| < \delta = 2^{-\tilde{\mathcal{O}}(n^3 \tau)}, \tag{4.11}$$

where $\delta$ is a certain threshold $|f(x_0, y_0)| + |g(x_0, y_0)|$ has to undercut. In the case where $\mathcal{B}$ does not contain a real solution of (1.1), Theorem 1 and (4.11) ensure that, for all $(x_0, y_0) \in B(\alpha, \beta)$, we have:

$$|f(x_0, y_0)| + |g(x_0, y_0)| \geq 2^{-\tilde{\mathcal{O}}(n^3 \tau)}. \tag{4.12}$$

Now, in order to certify or discard $(\alpha, \beta)$ as a solution of $f = g = 0$, we set $\rho := \lceil -\log s \rceil$ that is directly related to the size $s$ of $\mathcal{B}$ and evaluate $f(\alpha, \beta)$ and $g(\alpha, \beta)$ by means of interval arithmetic with precision $\rho$; see [14, Section 4] and Appendix B for the definition of *polynomial interval evaluation with precision $\rho$*. Obviously, the so obtained intervals $\mathfrak{B}(f(\alpha, \beta), \rho)$ and $\mathfrak{B}(g(\alpha, \beta), \rho)$ then contain $f(\mathcal{B})$ and $g(\mathcal{B})$ because each $(x_0, y_0) \in B$ approximates $(\alpha, \beta)$ to an error $2^{-\rho}$ or less. Now, if one of the intervals $\mathfrak{B}(f(\alpha, \beta), \rho)$ or $\mathfrak{B}(g(\alpha, \beta), \rho)$ does not contain zero, then $(\alpha, \beta)$ cannot be solution of (1.1). If, for all $(\epsilon_1, \epsilon_2) \in \mathfrak{B}(f(\alpha, \beta), \rho) \times \mathfrak{B}(g(\alpha, \beta), \rho)$, both inequalities $UB(\alpha, \beta, u^{(y)})|\epsilon_1| + UB(\alpha, \beta, v^{(y)})|\epsilon_2| < LB(\alpha)$ and $UB(\alpha, \beta, u^{(x)})|\epsilon_1| + UB(\alpha, \beta, v^{(x)})|\epsilon_2| < LB(\beta)$ are fulfilled, then $(\alpha, \beta)$ must be a solution. In the case where we cannot discard or certify $(\alpha, \beta)$ as

a solution, we refine $\mathcal{B}$ by refining the corresponding isolating intervals $I(\alpha)$ and $I(\beta)$ and proceed. According to (4.11) and (4.12), we eventually succeed if $\mathfrak{B}(f(\alpha,\beta),\rho)$ and $\mathfrak{B}(g(\alpha,\beta),\rho)$ have width less than some threshold larger than $2^{-\tilde{\mathcal{O}}(n^3\tau)}$. In the following consideration, we will bound the cost for the evaluation of $\mathfrak{B}(f(\alpha,\beta),\rho)$ and $\mathfrak{B}(f(\alpha,\beta),\rho)$ and for the refinement of $\mathcal{B}$.

*Complexity of polynomial evaluation.* In the previous section, we have shown that the candidate boxes $\mathcal{B} = B(\alpha,\beta)$ have to be refined to a size $s$ such that $\mathfrak{B}(f(\alpha,\beta),\rho)$ and $\mathfrak{B}(g(\alpha,\beta),\rho)$ have width less than a certain threshold larger than $2^{-\tilde{\mathcal{O}}(n^3\tau)}$, where we consider interval arithmetic with precision $\rho = \lceil -\log s \rceil$. We first estimate the absolute error induced by the interval arithmetic. Then, we estimate the overall cost for the polynomial evaluations at all candidate solutions.

**Theorem 4.** *The cost for evaluating $\mathfrak{B}(f(\alpha,\beta),\rho)$ and $\mathfrak{B}(g(\alpha,\beta),\rho)$ for all candidate solutions $(\alpha,\beta) \in \mathcal{C}$ using interval arithmetic with precision $\rho \in \mathbb{N}$ is bounded by $\tilde{\mathcal{O}}(n^5(\rho + n^2\tau))$ bit operations. Each of the so obtained intervals has width less than*
$$2^{-\rho}2^\tau B^{\mathcal{O}(n)} = 2^{-\rho+\tilde{\mathcal{O}}(n^2\tau)}.$$

*Proof.* Remark that evaluating a polynomial $F$ of degree $d$ at a point $x_0$ using fixed point arithmetic with precision $\rho$ demands for $\tilde{\mathcal{O}}(d(\rho + \log(\max\{|x_0|,1\})))$ bit operations. Since each root $\alpha$ of the resultant polynomial $R^{(y)}$ has modulus less than $B = 2^{\tilde{\mathcal{O}}(n\tau)}$, we can compute the coefficients $f_i(\alpha)$ of $f(\alpha,y)$ (by means of interval arithmetic with precision $\rho$) using $\tilde{\mathcal{O}}(n^2(\rho + n\log B))$ bit operations, while the resulting intervals contain values of modulus less than $2^\tau B^{\mathcal{O}(n)}$. Thus, the cost for evaluating all fiber polynomials $f(\alpha,y)$ is bounded by $\tilde{\mathcal{O}}(n^4(\rho + n\log B))$ because $R^{(y)}$ has at most $n^2$ many roots. By [14, Lem. 3], the absolute error for each coefficient $f_i(\alpha)$ is given by

$$|f_i(\alpha) - \mathfrak{B}(f_i(\alpha),\rho)| \leq 2^{-\rho+1}(n+1)^2 2^\tau B^n = 2^{-\rho}2^\tau B^{\mathcal{O}(n)}. \tag{4.13}$$

In a second step, we compute $\mathfrak{B}(f(\alpha,\beta),\rho)$ for a fixed $\alpha$. Again, for each $\beta$, we have to perform $n$ multiplications and additions involving numbers of bit length $\mathcal{O}(\rho + \tau + n\log B)$, hence, the cost for these evaluations is bounded by $\tilde{\mathcal{O}}(n(\rho+\tau+n\log B))$ bit operations. Summing up over all $\beta$ above a fixed $\alpha$ yields $\tilde{\mathcal{O}}(n^3(\rho+\tau+n\log B))$ bit operations and, thus, $\tilde{\mathcal{O}}(n^5(\rho+\tau+n\log B)) = \tilde{\mathcal{O}}(n^5(\rho+n^2\tau))$ constitutes an upper bound on the number of bit operations needed for all evaluations. The absolute error for each $f(\alpha,\beta)$ is bounded by $2^{-\rho}2^\tau B^{\mathcal{O}(n)}$ because each coefficient $f_i(\alpha)$ is approximated to an error $2^{-\rho}2^\tau B^{\mathcal{O}(n)}$ and the absolute value of each $\beta$ is bounded by $B$. The corresponding bounds for the evaluation of $\mathfrak{B}(g(\alpha,\beta),\rho)$ follow in completely analogous manner.

Now, according to the above Theorem and our considerations in the previous section, we must succeed for a precision $\rho \geq \rho_0 = \tilde{\mathcal{O}}(n^3\tau)$. Namely, since the width of the intervals $\mathfrak{B}(f(\alpha,\beta),\rho)$ and $\mathfrak{B}(g(\alpha,\beta),\rho)$ is bounded by $2^{-\rho+\tilde{\mathcal{O}}(n^2\tau)}$, it suffices to consider a precision $\rho_0 = \tilde{\mathcal{O}}(n^3\tau)$ to guarantee that the so obtained intervals have width less than a threshold that is lower bounded by $2^{-\tilde{\mathcal{O}}(n^3\tau)}$. Hence, the cost for all polynomial interval evaluations is bounded by

$$\tilde{\mathcal{O}}(n^5(n^3\tau + n^2\tau)) = \tilde{\mathcal{O}}(n^8\tau).$$

*Complexity of the root refinement.* We have already shown in the previous section that it suffices to refine each of the candidate boxes $B(\alpha,\beta) = I(\alpha) \times I(\beta)$ to a size $2^{-\tilde{\mathcal{O}}(n^3\tau)}$. As mentioned in Section 3, we use the Quadratic interval refinement (QIR) method to refine the isolating intervals $I(\alpha)$ and $I(\beta)$ for the real roots $\alpha$, $\beta$ of $R^{(y)}$ and $R^{(x)}$, respectively. Recent work [14] improves on the asymptotic complexity for this root refinement by considering approximations of polynomial coefficients to a certain precision. It is shown that, for a square-free polynomial $F$ with magnitude $(d,\mu)$ and root bound $\Gamma$, the algorithm refines isolating intervals for all real roots of $F$ to a width $2^{-L}$ or less using

$$\tilde{\mathcal{O}}(d(d\log\Gamma(F) + \Sigma(F))(\mu + d\log\Gamma(F) + \Sigma(F)) + d^2 L) \tag{4.14}$$

bit operations, see [14, Thm. 22]. We need to refine the real roots of the square-free part $R^*$ of $R = R^{(y)}$ to a width bounded by $2^{-\tilde{\mathcal{O}}(n^3\tau)}$. Using the above result, we can do so with

$$\tilde{\mathcal{O}}(n^2 \cdot n^3\tau \cdot (n(n+\tau) + n^3\tau) + n^4 \cdot n^3\tau) = \tilde{\mathcal{O}}(n^8\tau^2) \tag{4.15}$$

bit operations, where we used that $\Sigma(R^*) = \tilde{\mathcal{O}}(n^3\tau)$ (see (4.3)) and all roots of $R^{(y)}$ have modulus less than $B = 2^{\mathcal{O}(n(\log n+\tau))}$. Again, the corresponding bounds for refining the roots of $R^{(x)}$ follow in complete analogous manner.

## 5 Conclusions

We have derived a bound on the bit complexity of the algorithm BISOLVE to isolate the real solutions of a bivariate polynomial system. To the best of our knowledge, the derived complexity bound improves considerably upon the best known complexity bounds to date. Still, we suppose that the derived bounds are not tight and there is room for improvement. This mainly stems from the fact that the number of solutions of a bivariate system is in most cases largely overestimated while, at the same time, this number is intimately bound up with the root separation and coefficient bitlength of the resultant. We find it, therefore, reasonable to invest some research in the theory of *sparse resultants* where the geometry of exponent vectors of input polynomials is analyzed to obtain sharp estimates. A bottleneck in the current analysis is the isolation of the resultant polynomial which is assumed to be a general polynomial of magnitude $(n^2, n(\tau + \log n))$. Hence, the question arises whether isolating the roots of a resultant is possibly easier than of a general polynomial? Finally, it might be possible to improve the complexity of the validation step by considering asymptotically fast algorithms for multipoint polynomial evaluation [15].

## A  Proof of Theorem 3

We begin with the following auxiliary lemma.

**Lemma 3 (Davenport-Mahler).** *Let $F(x)$ be a polynomial with integer coefficients of magnitude $(d, \mu)$, and $V$ be the set of* distinct *complex roots $z_1, \ldots, z_r$ of $F$ ($r \leq d$). If a directed graph $\mathcal{G} = (V, E)$ satisfies conditions: **(a)** for every edge $(\alpha, \beta) \in E$, $|\alpha| \leq |\beta|$; **(b)** $\mathcal{G}$ is acyclic; and **(c)** the in-degree of any node is at most 1, then*

$$\prod_{(\alpha,\beta) \in E} |\alpha - \beta| \geq \frac{1}{(\sqrt{d+1}2^\mu)^{d-1}} \cdot \left(\frac{\sqrt{3}}{d}\right)^{\#E} \cdot \left(\frac{1}{d}\right)^{d/2}.$$

For a proof of the above lemma, we refer to [9, Corollary 3.11]. It exploits the fact that we can always factorize $F$ over $\mathbb{Z}$ and extract its square-free part $F^* \in \mathbb{Z}[x]$. Then, the generalized Davenport-Mahler bound applied to $F^*$ automatically extends to $F$ because for $F^*$, as the divisor of $F$, it holds that $\mathcal{M}(F^*) < \mathcal{M}(F)$.[3] Now we are ready to show the main result.

*Proof of Theorem 3.* We begin with clustering the roots of $F$ into subsets consisting of nearby roots. Without loss of generality, we assume that $\sigma(z_1, F) \leq \cdots \leq \sigma(z_r, F)$. For $h \in \mathbb{N}$, we denote $i(h)$ the maximal index $i$ with $\sigma(z_i, F) \leq 2^{-h}$ and $U = U(h) := \{z_1, \ldots, z_{i(h)}\}$ the corresponding set of roots $z_i$ with $\sigma(z_i, F) \leq 2^{-h}$. If $h \leq \log(1/\sigma(F))$, then $U$ contains at least two roots. Our goal is to partition $U$ into disjoint subsets $U_1, \ldots, U_l$ that contain closely located roots. It can be shown that for $h \leq \log(1/\sigma(F))$, there exists such a partition of $U$ that $|U_i| \geq 2$ for all $i = 1, \ldots, l$ and $|z - z'| \leq d2^{-h}$ for all $z, z' \in U_i$, see [17, Lem. 19].

We consider a directed graph $\mathcal{G}_i$ on each $U_i$ connecting consecutive roots of $U_i$ in ascending order of their absolute values. Let $\mathcal{G} := (U, E)$ be the union of all $\mathcal{G}_i$. Remark that, $\mathcal{G}$ is a directed graph on $U$, and the conditions of Lemma 3 are satisfied. Moreover, since each set $U_i$ contains at least two roots, we must have $\#E \geq i(h)/2$. Additionally, it holds that $\#E < i(h)$, where in corner case all roots of $F$ belong to a single partition $U_1$. Hence, by Lemma 3:

$$(d2^{-h})^{\frac{i(h)}{2}} > \prod_{(\alpha,\beta) \in E} |\alpha - \beta| \geq \frac{1}{(\sqrt{d+1}2^\mu)^{d-1}} \cdot \left(\frac{\sqrt{3}}{d}\right)^{i(h)} \cdot \left(\frac{1}{d}\right)^{\frac{d}{2}}$$

$$> \frac{1}{(d+1)^d 2^{d\mu}} \cdot \left(\frac{3}{d^2}\right)^{\frac{i(h)}{2}},$$

and therefore,

$$i(h) < \frac{2d(\mu + \log(d+1))}{\log 3 + \log d + h} < \frac{2d(\mu + \log(d+1))}{h}.$$

---

[3] Here, $\mathcal{M}(F)$ denotes the Mahler measure for $F$.

By combining the above inequality with $h \leq \log(1/\sigma(F))$, and $i(h) \geq 2$ (because $|U_i| \geq 2$), we conclude that: $\log(1/\sigma(F)) < d(\mu + \log(d+1)) + 1$. Otherwise, there would exist an $h$ with $i(h) < 2$ which is not possible. For the bound on $\Sigma(F)$, it suffices to consider only the roots $z_1, \ldots, z_k$ with separation $\leq 1/2$ since all other roots contribute with at most $d$ to the sum $\Sigma(F)$. Let $h_{\max} = \lceil d(\mu + \log(d+1)) \rceil$, then:

$$\Sigma(F) = \sum_{i=1}^{k} \log(1/\sigma(z_i, F)) < \sum_{h=1}^{h_{\max}} i(h) < d(\mu + \log(d+1)) \sum_{h=1}^{h_{\max}} \frac{1}{h} = O(d\mu \log(d\mu)).$$

## B Approximate polynomial evaluation

In this section we define some basic operations on approximate numbers and establish the error bounds for polynomial evaluation, we refer to [14, Section 4] for comprehensive discussion. For $\rho \in \mathbb{N}$ and $x \in \mathbb{R}$, we define:

$$\text{down}(x, \rho) = \{k \cdot 2^{-\rho} \in \mathbb{R} | k = \lfloor x \cdot 2^{\rho} \rfloor\}, \text{ and } \text{up}(x, \rho) = \{k \cdot 2^{-\rho} \in \mathbb{R} | k = \lceil x \cdot 2^{\rho} \rceil\}.$$

Meaning that, $x$ is included in an interval: $\mathfrak{B}(x, \rho) := [\text{down}(x, \rho), \text{up}(x, \rho)]$. In what follows, we will omit $\rho$ and write $\text{up}(x)$ or $\mathfrak{B}(x)$ to simplify the notation. Arithmetic operations on approximate numbers obey the rules of classical interval arithmetic. For two numbers $x, y \in \mathbb{R}$, we define:

$$\mathfrak{B}(x) + \mathfrak{B}(y) := [\text{down}(x) + \text{down}(y), \text{up}(x) + \text{up}(y)],$$
$$\mathfrak{B}(x) - \mathfrak{B}(y) := [\text{down}(x) - \text{up}(y), \text{up}(x) - \text{down}(y)],$$
$$\mathfrak{B}(x) \cdot \mathfrak{B}(y) := \left[ \min_{i,j=\{1,2\}} \{H_i(x) H_j(y)\}, \max_{i,j=\{1,2\}} \{H_i(x) H_j(y)\} \right],$$
$$\text{with } H_1(x) = \text{down}(x), \text{ and } H_2(x) = \text{up}(x).$$

Using the above rules, for a polynomial $F$ and $z \in \mathbb{R}$, $\mathfrak{B}(F(z), \rho)$ can be defined by expanding the polynomial according to Horner scheme:

$$\mathfrak{B}(F(z)) = \mathfrak{B}(F_0) + \mathfrak{B}(z) \cdot (\mathfrak{B}(F_1) + \mathfrak{B}(z) \cdot (\mathfrak{B}(F_2) + \ldots)).$$

The next lemma bounds the error of polynomial evaluation with precision $\rho$:

**Lemma 4.** *Let $F$ be a polynomial with magnitude $(d, \mu)$, $c \in \mathbb{R}$ with $|c| \leq 2^{\upsilon}$, and $\rho \in \mathbb{N}$. Then,*

$$|F(c) - H(F(c), \rho)| \leq 2^{-\rho+1}(d+1)^2 2^{\mu+d\upsilon},$$

*where $H = \{\text{down}, \text{up}\}$. In particular, $\mathfrak{B}(F(c), \rho)$ has width $2^{-\rho+2}(d+1)^2 2^{\mu+d\upsilon}$ or less. For a proof, see [14, Lem. 3].*

In essence, the lemma asserts that the absolute error result from approximate polynomial evaluation is *linear* in $2^{-\rho}$, which has important consequences for us in Section 4.3.